\documentclass[amsmath,jcp,reprint,twocolumn]{revtex4-1}

\usepackage{amssymb}
\usepackage{graphicx}
\usepackage{color}

\begin{document}

\title{Simulation of optical response functions in molecular junctions}

\author{Yi Gao}
\affiliation{Department of Chemistry and Biochemistry, University of California San Diego, La Jolla, CA 92093, USA}
\author{Michael Galperin}
\affiliation{Department of Chemistry and Biochemistry, University of California San Diego, La Jolla, CA 92093, USA}

\begin{abstract}
We discuss theoretical approaches to nonlinear optical spectroscopy of molecular junctions. 
Optical response functions are derived in the form convenient for 
implementation of Green function techniques, and their expressions 
in terms of pseudoparticle nonequilibrium Green functions are proposed.
The formulation allows to account for both intra-molecular interactions and
hybridization of molecular states due to coupling to contacts.
Two-dimensional optical spectroscopy in junctions is considered as 
an example.  
\end{abstract}

\maketitle

%%%%%%%%%%%%%%%%%%%%%%%%%%%%%%%%%%%%%%%%%%%%%%
%%%%%%%%%%%%%%%%%%%%%%%%%%%%%%%%%%%%%%%%%%%%%%

\section{Introduction}\label{intro}
Interaction of light with matter is an established field of research 
providing spectroscopic tools for study of interactions and dynamical processes. 
In molecular systems nonlinear optical spectroscopy is utilized 
to study transient molecular phenomena~\cite{TianScience03,WortmannChemistry03,FlemingJCP04,FischerCHIRALITY05,OsukaKimJACS06,FlemingPNAS06,GoodsonAccChemRes05,ZhuScience11} and local interactions~\cite{FayerCPL04,SkinnerJCP05},
for driving~\cite{FennelRMP10} and coherent control~\cite{ShapiroBrumer_2003,GoswamiPhysRep03,SilberbergAnnRevPhysChem09},
and for molecular imaging~\cite{DavisRevSciInstr08,XieAnnRevPhysChem11}.
Theory of nonlinear optical spectroscopy was developed~\cite{Mukamel_1995,MukamelRMP98}  
and successfully utilized in numerous studies~\cite{MukamelPRA89,OkumuraJCP97,OkumuraTanimuraJCP97,TokmakoffJPCA00,OvchinnikovApkarianVothJCP01,FlemingJPCA01,MayPhysRep01,OkumuraJPCA03,MukamelChemRev04,MukamelPRA05,NeivandtApplSpecRev05,MukamelPRB08,MukamelPRB09}.

Rapid development of nanofabrication techniques made possible to study interaction of
light with molecular junctions~\cite{CheshnovskySelzerNatNano08,NatelsonNL08,NatelsonNatNano11,ApkarianACSNano12,ApkarianHessNL13} 
leading to appearance of molecular optoelectronics~\cite{MGANPCCP12}.
In molecular junctions electron participate in both optical scattering and quantum transport.
Theoretical challenge is description of the two processes on the same footing.
{\em Ab initio} simulations in the field of molecular electronics employ
combination of the nonequilibrium Green function (NEGF) method 
with density functional theory (DFT)~\cite{XueDattaRatnerCP02,TaylorStokbroPRB02}. 
Both NEGF and DFT are formulated in the language of quasiparticles (orbitals);
and current through the junction is the primary goal of simulations.
Thus it is natural that one of the approaches to describe optical spectroscopy of
molecular junctions utilizes  quasiparticle language with photon flux giving information 
on optical response of the system~\cite{GalperinNitzanJCP06,GalperinTretiakJCP08,GalperinRatnerNitzanNL09,GalperinRatnerNitzanJCP09,MGANJPCL11,MGANPRB11,ParkMGEPL11,ParkMGPRB11,OrenMGANPRB12,ParkMGPST12,BanikApkarianParkMGJPCL13,ApkarianMGANPRB16}. 
These formulations are capable to account for molecule-contacts coupling exactly,
while intra-molecular Interactions are usually treated perturbatively.

Traditional nonlinear optical spectroscopy is formulated in the language of many-body 
states of isolated molecule (or utilizing dressed states picture)~\cite{Mukamel_1995,Nitzan_2006}, which allows
to describe all the intra-molecular interactions exactly.
In molecular junctions the formulation is complemented by quantum master equation (QME)
to account for current carrying state of the system~\cite{MukamelJCP14,MukamelJCP15}.
Molecular spectroscopy is characterized via response functions
obtained from perturbative expansion of photon flux in molecular interaction
with radiation field, which requires evaluation of multi-time correlation functions.
The latter are usually calculated employing quantum regression theorem~\cite{BreuerPetruccione_2003}.
Within the approach interactions with the field define time intervals
in which evolution of the system is governed by reduced Liouvillian,
while every interaction with optical field also implies destruction of the
molecule-contacts entanglement.
The latter is an artifact of the formulation, which (as we show below) may be problematic. 

A possible alternative to the QME is utilization of Green function
methods of the nonequilibrium atomic limit formulations~\cite{WhiteOchoaMGJPCC14}.
These formulations provide consistent way of taking into account molecule-contacts 
coupling and keep system-bath entanglement intact while the system interact with 
radiation field. Recently, we utilized the approach to
generalize our previous quasiparticle (orbital) based formulations for Raman spectroscopy 
in molecular junctions~\cite{GalperinRatnerNitzanNL09,GalperinRatnerNitzanJCP09,MGANJPCL11,MGANPRB11}
to account exactly for dependence of molecular normal modes on charging state 
of the molecule~\cite{WhiteTretiakNL14}.
Here we utilize the nonequilibrium atomic limit tools to complement traditional 
nonlinear optical spectroscopy formulations~\cite{MukamelJCP14,MukamelJCP15}. 

 Structure of the paper is the following. After introducing model of molecular junction
 in Section~\ref{model} we discuss a derivation of optical response functions
 in  form convenient for implementation of Green function techniques
 (Section~\ref{respfunc}). Section~\ref{pprespfunc} introduces the 
 pseudoparticle nonequilibrium Green functions (PP-NEGF) formulation
 for the response functions of nonlinear optical spectroscopy.
 In Section~\ref{coh2dsig} we specialize our formulation to 
 description of coherent multi-dimensional optical signals in junctions
 following recent consideration in Ref.~\cite{MukamelJCP15}.
 Section~\ref{conclude} concludes.
 
 %%%%%%%%%%%%%%%%%%%%%%%%%%%%%%%%%%%%%%%%%%%%%%
\begin{figure}[htbp]
\centering\includegraphics[width=\linewidth]{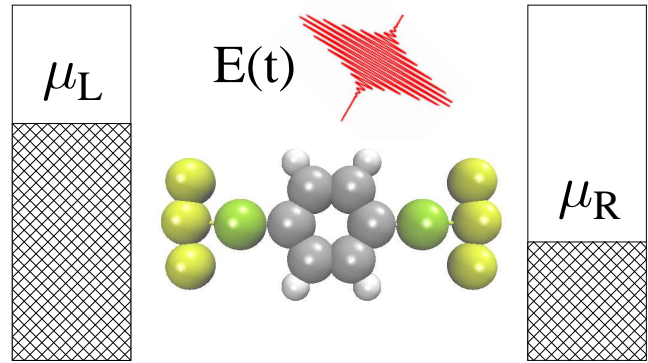}
\caption{\label{fig1}
Sketch of a molecular junction subjected to external radiation field.
}
\end{figure}
%%%%%%%%%%%%%%%%%%%%%%%%%%%%%%%%%%%%%%%%%%%%%%

%%%%%%%%%%%%%%%%%%%%%%%%%%%%%%%%%%%%%%%%%%%%%%

\section{Model}\label{model}
We consider model of a junction consisting of a single molecule $M$ coupled to two metallic
contacts $L$ and $R$ and subjected to external radiation field $F$ (see Fig.~\ref{fig1}). 
The contacts are electron reservoirs each at it own equilibrium characterized by 
electrochemical potentials $\mu_{L,R}$ and temperatures $T_{L,R}$.  
The field will be treated semi-classically. Following Ref.~\cite{Mukamel_1995}
we derive expression for the optical signal assuming quantum radiation field
and transfer to classical description when modeling molecule-field interaction
(see details below).
In accordance with common practice of molecular spectroscopy formulations
below we utilize many-body states $\lvert S\rangle$ of the isolated molecule as a basis.
Depending on particular problem these may be electronic or vibronic states of the molecule.
We assume that coupling to radiation field is restricted to molecular subspace.   
Hamiltonian of the model is
\begin{equation}
\label{H}
 \hat H=\hat H_M+\sum_{K=L,R}\left(\hat H_K+\hat V_{MK}\right)
 + \hat H_F + \hat V_{MF}
\end{equation} 
Here $\hat H_M$, $\hat H_K$ ($K=L,R$) and $\hat H_F$ represent
molecule, contacts and radiation field, respectively. 
$\hat V_{MK}$ and $\hat V_{MF}$ describe molecular coupling to the contacts and field.
Explicit expressions are (here and below $e=\hbar=k_B=1$)
\begin{align}
\label{HM}
\hat H_M =& \sum_{S\in M} E_S\hat X_{SS}
\\
\label{HK}
\hat H_K =& \sum_{k\in K} \varepsilon_k\hat c_k^\dagger\hat c_k
\\
\label{VMK}
\hat V_{MK}=& \sum_{k\in K}\sum_{S_1,S_2\in M}
\left( V_{k,S_1S_2}\hat c_k^\dagger \hat X_{S_1S_2} + H.c. \right)
\\
\label{HF}
\hat H_F =& \sum_{\alpha}\omega_\alpha\hat a_\alpha^\dagger\hat a_\alpha
\\
\label{VMF}
\hat V_{MF} =& -\sum_\alpha\sum_{S_1,S_2\in M} 
\left( \hat {\mathcal{E}}^\dagger(t)\mu_{S_1S_2}\hat X_{S_1S_2} + H.c. \right) 
\end{align}
Here $\hat c_k^\dagger$ ($\hat c_k$) and $\hat a_\alpha^\dagger$ ($\hat a_\alpha$)
create (annihilate) respectively electron in level $k$ of the contact 
and photon in mode $\alpha$ of the field, 
$\hat X_{S_1S_2}\equiv \lvert S_1\rangle\langle S_2\rvert$ is the Hubbard operator,
and $\mu_{S_1S_2}$ is matrix element of molecular dipole moment operator.
$\hat V_{MF}$ is written in the rotating wave approximation and 
\begin{equation}
\hat {\mathcal{E}} = \sum_\alpha i\bigg(\frac{2\pi\omega_\alpha}{V}\bigg)^{1/2} \hat a_\alpha %e^{-i\omega_\alpha t} 
\end{equation}
is the positive frequency component of the field written in the long wavelength approximation, which we treat quantum mechanically.

%%%%%%%%%%%%%%%%%%%%%%%%%%%%%%%%%%%%%%%%%%%%%%

%%%%%%%%%%%%%%%%%%%%%%%%%%%%%%%%%%%%%%%%%%%%%%
\begin{figure*}[htbp]
\centering\includegraphics[width=\linewidth]{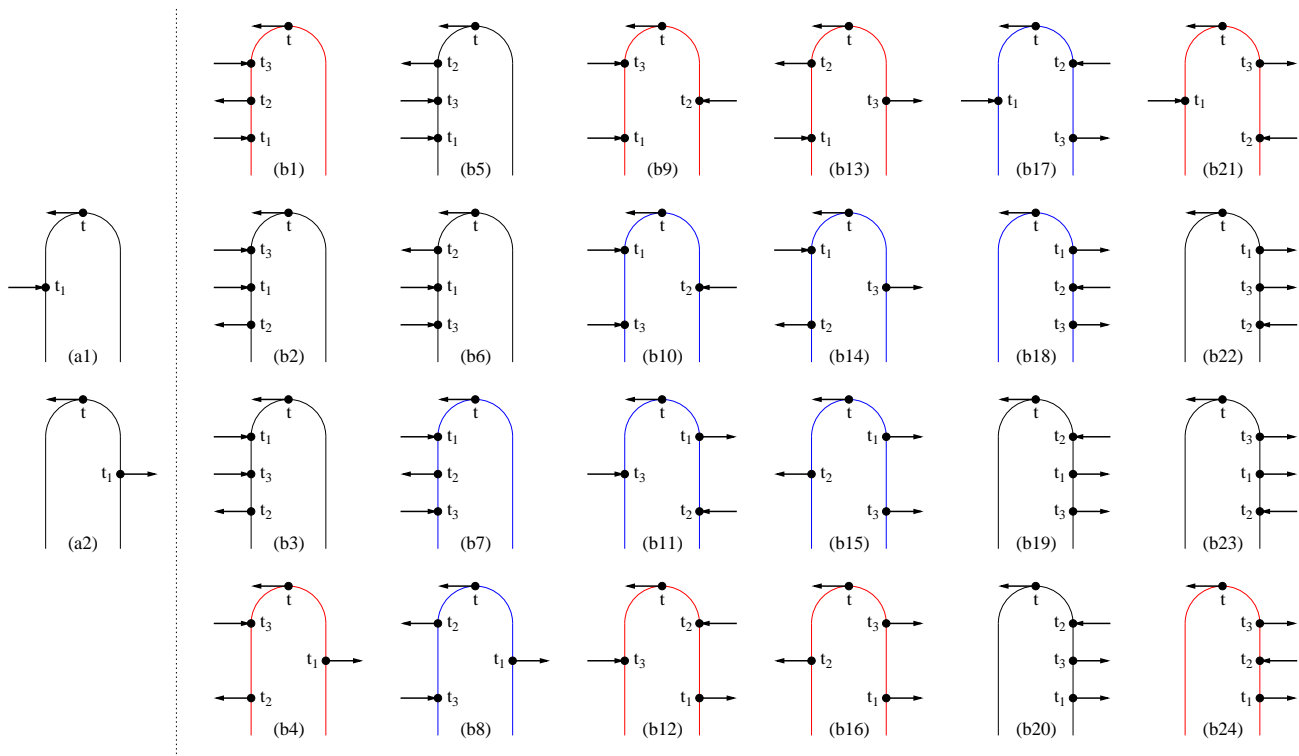}
\caption{\label{fig2}
All possible projections of contour variables $\tau_i$ for (a) linear, Eq.(\ref{S1c}),
and (b) third order, Eq.(\ref{S3c}), response functions. $t_i$ are physical times
corresponding to contour variables $\tau_i$. Lines pointing to right correspond to
positive frequency component of the field, Eq.(\ref{Ecl}); lines pointing to left
indicate their complex conjugate (negative component of the field).
}
\end{figure*}
%%%%%%%%%%%%%%%%%%%%%%%%%%%%%%%%%%%%%%%%%%%%%%

\subsection{Optical response functions}\label{respfunc}
Following Ref.~\cite{Mukamel_1995} we define optical signal (photon flux) 
as rate of change of population of the radiation field modes
\begin{equation}
\label{Sdef}
 S(t) \equiv \frac{d}{dt} \sum_\alpha \langle \hat a_\alpha^\dagger(t)\hat a_\alpha(t)\rangle
\end{equation}
where $\langle\ldots\rangle\equiv \mbox{Tr}[\ldots\hat \rho]$ indicates quantum and 
statistical average with density operator $\hat \rho$ of the whole system,
operators $\hat a^\dagger(t)$ and $\hat a(t)$ are in the Heisenberg picture. 
Note that the field is treated quantum-mechanically in (\ref{Sdef}). 
Utilizing Heisenberg equation of motion one gets for the model (\ref{H})-(\ref{VMF})
\begin{equation}
 \label{S}
 S(t) = -2\,\mbox{Im} \sum_{S_1,S_2}\mu_{S_1S_2}
 \langle\hat {\mathcal{E}}^\dagger(t)\hat X_{S_1S_2}(t)\rangle
\end{equation}
Further treatment relies on perturbative expansion of (\ref{S}) in the
molecule-field coupling $\hat V_{MF}$~\cite{Mukamel_1995}. 
The expansion is performed on the Keldysh contour~\cite{HaugJauho_2008} 
to account for nonequilibrium character of the molecular junction.
Expressing Eq.(\ref{S}) in interaction picture and expanding resulting scattering 
operator in Taylor series leads to
\begin{equation}
\label{SPT}
S(t) = \sum_{n=0}^{\infty} S^{(2n+1)}(t)
\end{equation}
where
\begin{align}
\label{SnPT}
&S^{(m)}(t) = -2\,\mbox{Im}\frac{(-i)^m}{m!}\sum_{S_1,S_2}\mu_{S_1S_2}
\int_c d\tau_1 \ldots \int_c d\tau_m 
\nonumber \\ &\qquad
\langle T_c\,
\hat {\mathcal{E}}^\dagger(t)\hat X_{S_1S_2}(t)\, \hat V_{MF}(\tau_1) \ldots
\hat V_{MF}(\tau_m)\rangle_0
\end{align}
Here $\tau_i$ ($i=1,\ldots,m$) are contour variables, $T_c$ is the contour ordering 
operator, operators are in the interaction picture, and subscript $0$ indicates
evolution under zero-order Hamiltonian 
$\hat H_0\equiv \hat H_M+\sum_{K=L,R}\left(\hat H_K+\hat V_{MK}\right)
 + \hat H_F$. 
Note that even contributions in the expansion (\ref{SPT}) drop out due to 
odd number of photon creation/annihilation operators in the correlation 
function~\footnote{For classical field these terms drop out by symmetry in the case of 
isotropic medium~\cite{Mukamel_1995}.}.
Substituting explicit expression for $\hat V_{MF}$ into  (\ref{SnPT}) 
leads to explicit expressions. In particular,
the first ($m=1$) and second ($m=3$) contributions in the expansion (\ref{SPT}),
which represent respectively linear and third order response, are
\begin{align}
\label{S1c}
& S^{(1)}(t) = -2\,\mbox{Re}\sum_{\substack{S_1^{(0)},S_2^{(0)} \\ S_1^{(1)},S_2^{(1)}}}
\mu_{S_1^{(0)}S_2^{(0)}}\mu_{S_1^{(1)}S_2^{(1)}}^{*}
\\ & \qquad
\int_c d\tau_1
\langle T_c \, \hat {\mathcal{E}}^\dagger(t)\hat X_{S_1^{(0)}S_2^{(0)}}(t) \,
\hat X_{S_1^{(1)}S_2^{(1)}}^\dagger(\tau_1)\hat {\mathcal{E}}^\dagger(\tau_1)
\rangle_0
\nonumber
\\
\label{S3c}
&S^{(3)}(t) = \mbox{Re} \sum_{\substack{S_1^{(0)},S_2^{(0)},S_1^{(1)},S_2^{(1)}\\
                                                                   S_1^{(2)},S_2^{(2)},S_1^{(3)},S_2^{(3)}}}
\mu_{S_1^{(0)}S_2^{(0)}}\mu_{S_1^{(1)}S_2^{(1)}}^{*}
\nonumber \\ &\qquad
\mu_{S_1^{(2)}S_2^{(2)}}\mu_{S_1^{(3)}S_2^{(3)}}^{*}
\int_c d\tau_1\int_c d\tau_2\int_c d\tau_3
\\ &\qquad
\langle T_c \, \hat {\mathcal{E}}^\dagger(t)\hat X_{S_1^{(0)}S_2^{(0)}}(t) \,
\hat X_{S_1^{(1)}S_2^{(1)}}^\dagger(\tau_1)\hat {\mathcal{E}}(\tau_1)\,
\nonumber \\ & \qquad \times
\hat {\mathcal{E}}^\dagger(\tau_2)\hat X_{S_1^{(2)}S_2^{(2)}}(\tau_2) \,
\hat X_{S_1^{(3)}S_2^{(3)}}^\dagger(\tau_3)\hat {\mathcal{E}}(\tau_3)
\rangle_0
\nonumber
\end{align}
All possible placements (projections) of contour variables $\tau_i$
relative to time of the signal $t$ and to each other 
are shown in Fig.~\ref{fig2}a for the linear, Eq.(\ref{S1c}), and Fig.~\ref{fig2}b for 
the third order, Eq.(\ref{S3c}), response functions. These projections are
the double sided Feynman diagrams~\cite{Mukamel_1995}. Note difference
between Hilbert (shown in Fig.~\ref{fig2}) and Liouville space projections with the former
keeping ordering along the contour while the latter imposing additional restrictions
on ordering along the real time axis. As a result number of Liouville space projections
is bigger. 

Following Ref.~\cite{MukamelPRA08} at this point we assume that the incoming field 
is in a coherent state and transfer to classical representation\footnote{Note that at this point 
we lose photon induced electron electron interaction.}
\begin{equation}
\label{Ecl}
 \mathcal{E}(t) = \sum_\alpha E_\alpha(t) e^{-i\omega_\alpha t + i\phi_\alpha}
\end{equation} 
where $E_\alpha(t)$ is complex time dependent envelope representing, e.g., laser pulse.
Thus expression for optical response only requires evaluation of electronic
multi-time correlation function.

%%%%%%%%%%%%%%%%%%%%%%%%%%%%%%%%%%%%%%%%%%%%%%

\subsection{Pseudoparticle formulation}\label{pprespfunc}
Evaluation of electronic multi-time correlation functions of the type present in 
Eqs.~(\ref{SnPT})-(\ref{S3c}) is a complicated problem,
which may be approximately treated with a number of techniques.  
Quantum regression theorem~\cite{BreuerPetruccione_2003} is the usual
choice in nonlinear optical spectroscopy~\cite{Mukamel_1995}.
For example, recent works on optical spectroscopy in junctions~\cite{MukamelJCP14,MukamelJCP15}
utilize this approach.
Within the approach interactions with the field define time intervals
in which evolution of the system is governed by reduced Liouvillian.
The approach destroys correlations between molecule
and contacts at every instant of interaction with the field.  
Below we demonstrate that the approximation may be problematic.

We utilize the pseudoparticle nonequilibrium Green function (PP-NEGF) methodology~\cite{EcksteinWernerPRB10,OhAhnBubanjaPRB11,WhiteGalperinPCCP12,WernerRMP14}
as an alternative formulation, which describes molecular system
utilizing many-body states, accounts for molecular hybridization due to coupling to contacts,
and avoids assumption of molecule-contacts destruction of coherence
at times of interaction with radiation field. 
PP-NEGF has several important advantages:
1.~It is conceptually simple; 2.~Its practical implementations rely on a set
of controlled approximations; 3.~Already at the lowest order of the theory,
the non-crossing approximation (NCA), it goes beyond usual QME 
approaches by accounting for both non-Markov effects and hybridization
of molecular states; 4.~the method is capable of treating the system in the basis of its
many-body states.  
Recently we applied the PP-NEGF to describe results of Raman scattering
experiment  in the OPV3 molecular junction~\cite{WhiteTretiakNL14}. 
Here we utilize it for a more general description of optical response functions 
in expansion (\ref{SPT}). 

 PP-NEGF introduces second quantization in the space of many-body states 
 $\lvert S\rangle$ of a system. Pseudoparticle operators $\hat p_S^\dagger$ ($\hat p_S$)
 create (annihilate) state $\lvert S\rangle$
 \begin{equation}
  \hat p_S^\dagger \lvert vac\rangle = \lvert S\rangle
 \end{equation}
 where $\lvert vac\rangle$ is vacuum state. Thus Hubbard operators,
 which appear in the response functions, Eqs.~(\ref{S}), (\ref{S1c})
 and (\ref{S3c}), can be expressed as
 \begin{equation}
  \hat X_{S_1S_2} \equiv\lvert S_1\rangle\langle S_2\rvert
  \rightarrow \hat p_{S_1}^\dagger\hat p_{S_2}
 \end{equation}
The consideration requires extended Hilbert space formulation, physical subspace
of which is defined by the normalization condition
\begin{equation}
 \label{norm}
 \sum_S \hat p_S^\dagger\hat p_S = 1 
\end{equation} 
In the extended Hilbert space the formulation utilizes standard tools of 
the quantum field theory. Restriction (\ref{norm}) modifies resulting expressions 
projecting them onto the physical subspace.

Evaluation of multi-time correlation functions is complicated by the non-quadratic character 
of the molecule-contacts coupling, Eq.(\ref{VMK}), in the pseudoparticle representation.
Usual perturbative treatment requires expanding correlation functions in the interaction
$\hat V_{MK}$ up to a particular order, identifying irreducible diagrams, dressing them 
and formulating corresponding equations-of-motion. 
Already at the level of linear response, Eq.(\ref{S1c}), this will require simultaneous
solution of the Dyson and Bethe-Salpeter equations. For simplicity here we rely
on a mean-field treatment, where a multi-time correlation function can be 
approximately presented as a product of pseudoparticle Green functions
\begin{equation}
 \label{Gdef}
 G_{S_1S_2}(\tau_1,\tau_2)\equiv 
  -i\langle T_c\,\hat p_{S_1}(\tau_1)\,\hat p_{S_2}^\dagger(\tau_2)\rangle
\end{equation}
Below we demonstrate that in physically relevant range of parameters 
the approximation yields reasonable description of optical response.  

Keeping in mind that the restriction (\ref{norm}) only allows one lesser 
pseudoparticle Green function to be present in any diagram~\cite{EcksteinWernerPRB10} 
after projection (see Fig.~\ref{fig2}) we get for an arbitrary correlation function  
the following approximate expression
\begin{align}
\label{cfapprox}
 &\langle\hat p_{1}^\dagger(t_1)\hat p_2(t_1)\hat p_3^\dagger(t_2)\hat p_4(t_2)\ldots
 \hat p_{2m-1}^\dagger(t_m)\hat p_{2m}(t_m)\rangle \approx
 \\ &
 i^m\zeta_1G^{<}_{2m,1}(t_m,t_1)G^{>}_{23}(t_1,t_2)\ldots G^{>}_{2m-2,2m-1}(t_{m-1},t_m)
 \nonumber
\end{align}
where $G^{<}$ ($G^{>}$) is lesser (greater) projection of the Green function (\ref{Gdef})
and $\zeta_1=-1$ ($+1$) if many-body state $1$ is of Fermi (Bose) type~\footnote{Note that all states of a optical correlation function are of the same type.}.

In the extended Hilbert space pseudoparticle Green functions, Eq.(\ref{Gdef}),
satisfy the usual Dyson equation. 
At steady state one has to solve equations for retarded and lesser projections~\cite{OhAhnBubanjaPRB11} 
\begin{align}
\label{GrEOM}
 \mathbf{G}^r(E)=&\bigg[E\mathbf{I}-\mathbf{H_M}-\mathbf{\Sigma}^r(E)\bigg]^{-1}
 \\
 \label{GltEOM}
 \mathbf{G}^{<}(E)=& \mathbf{G}^r(E)\,\mathbf{\Sigma}^{<}(E)\,\mathbf{G}^a(E)
\end{align}
Here Green functions $\mathbf{G}$, molecular Hamiltonian $\mathbf{H}_M$ 
and self-energies due to molecule-contacts coupling $\mathbf{\Sigma}$ 
are matricies in the basis of many-body states of an isolated molecule,
$\mathbf{I}$ is unity matrix, and $\mathbf{G}^a(E)=[\mathbf{G}^r(E)]^\dagger$ 
is advanced projection. 
Each of the expressions (\ref{GrEOM}) and (\ref{GltEOM}) have  to be solved
self-consistently, since within the formulation self-energies depend on Green functions.
The two equations belong to different subspaces and thus should be solved independently:
after (\ref{GrEOM}) converges its result (retarded projection of the Green function) 
is utilized in self-consistent solution of (\ref{GltEOM}).
Note that normalization (\ref{norm}) implies the following connection between 
greater and retarded projections of the Green function~\cite{WingreenMeirPRB94}
\begin{equation}
\mathbf{G}^{>}(E)=2\, i\, \mbox{Im}\, \mathbf{G}^r(E)
\end{equation}
For further details and explicit expressions for the self energies see, e.g., 
Ref.~\cite{WhiteGalperinPCCP12}.

%%%%%%%%%%%%%%%%%%%%%%%%%%%%%%%%%%%%%%%%%%%%%%

\section{Coherent 2D signals}\label{coh2dsig}
\begin{figure}[htpb]
\centering\includegraphics[width=\linewidth]{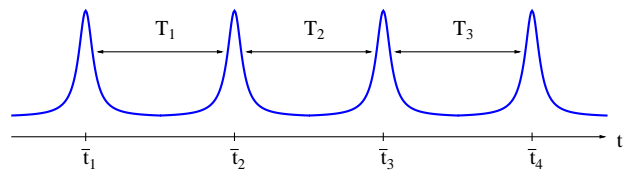}
\caption{\label{fig3}
Laser pulse sequence for time-domain experiment.
}
\end{figure}

Following consideration in Ref.~\cite{MukamelJCP15}
we now specify to 4-laser pulse sequence for time domain experiment.
Radiation field (\ref{Ecl}) takes the form
\begin{equation}
\label{Ecl4}
\mathcal{E}(t) = \sum_{i=1}^4
 E_i(t) e^{-i\omega_i t + i\phi_i}
\end{equation}
where $E_i(t)$ is the complex envelope of $i^{th}$ pulse centered around $\bar t_i$  
(see Fig.~\ref{fig3}). Following Ref.~\cite{MukamelJCP15} we will be interested in
stimulated signal in the fourth order of optical field with phase signature 
$\phi\equiv\phi_1-\phi_2+\phi_3-\phi_4$. 

%%%%%%%%%%%%%%%%%%%%%%%%%%%%%%%%%%%%%%%%%%%%%%
%%%%%%%%%%%%%%%%%%%%%%%%%%%%%%%%%%%%%%%%%%%%%%
%%%%%%%%%%%%%%%%%%%%%%%%%%%%%%%%%%%%%%%%%%%%%%
\begin{figure}[htpb]
\centering\includegraphics[width=\linewidth]{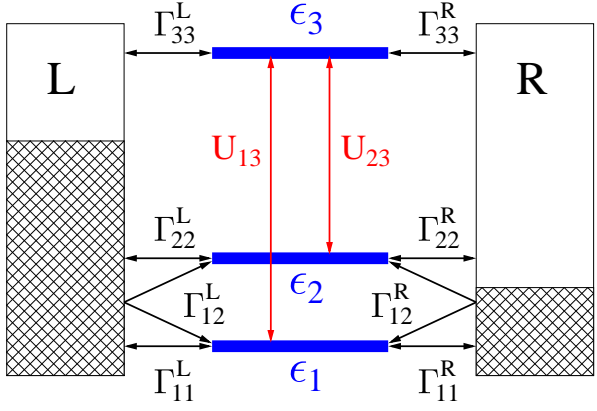}
\caption{\label{fig4}
A model of molecular junction.
}
\end{figure}
%%%%%%%%%%%%%%%%%%%%%%%%%%%%%%%%%%%%%%%%%%%%%%

The signal is is given by $8$ projections b1, b4, b9, b12, b13, b16, b21, b24
(and their analogs with $t_1\leftrightarrow t_3$: b7, b14, b10, b17, b8, b15, b11, b18)
of the third order response (\ref{S3c}) under restriction $t>t_3>t_2>t_1$
(compare with Fig.~2 of Ref.~\cite{MukamelJCP15}). Explicit expression is
\begin{align}
\label{S3t}
& S^{(3)}(t) = 2\,\mbox{Re}\sum_{\substack{S_1^{(0)},S_2^{(0)},S_1^{(1)},S_2^{(1)}\\
                                                                      S_1^{(2)},S_2^{(2)},S_1^{(3)},S_2^{(3)}}}
\int_{-\infty}^t dt_3 \int_{-\infty}^{t_3} dt_2 \int_{-\infty}^{t_2} dt_1
\nonumber \\ &
\mu_{S_1^{(0)}S_2^{(0)}}\mu_{S_2^{(1)}S_1^{(1)}}
\mu_{S_1^{(2)}S_2^{(2)}}\mu_{S_2^{(3)}S_1^{(3)}}
\mathcal{E}^{*}(t)\mathcal{E}(t_3)\mathcal{E}^{*}(t_2)\mathcal{E}(t_1)
\nonumber \\ &
\big\langle \big[\big[\big[
\hat p_{S_1^{(0)}}^\dagger(t)\hat p_{S_2^{(0)}}(t);
\hat p_{S_2^{(3)}}^\dagger(t_3)\hat p_{S_1^{(3)}}(t_3)\big];
\\ & \qquad
\hat p_{S_1^{(2)}}^\dagger(t_2)\hat p_{S_2^{(2)}}(t_2)\big];
\hat p_{S_2^{(1)}}^\dagger(t_1)\hat p_{S_1^{(1)}}(t_1)\big]
\big\rangle
\nonumber
\end{align}

We now introduce the total stimulated signal (see Fig.~\ref{fig3})
\begin{equation}
 S_{stim}(T_3,T_2,T_1)\equiv \int_{-\infty}^{+\infty} dt\, S^{(3)}(t)
\end{equation}
Assuming short impulses, $E_i(t)\approx E_i\delta(t-\bar t_i)$ and
utilizing approximation (\ref{cfapprox}) in (\ref{S3t}) 
we get for steady-state transport
\begin{widetext}
\begin{align}
\label{SstimOmT}
& S_{stim}(\Omega_1,T_2,\Omega_3)\equiv 
\int_0^\infty dT_3\int_0^\infty dT_1 e^{i\Omega_3T_3+i\Omega_1 T_1}
S_{stim}(T_3,T_2,T_1)
= 2\,\mbox{Re}\sum_{\substack{S_1^{(0)},S_2^{(0)},S_1^{(1)},S_2^{(1)}\\
                                                   S_1^{(2)},S_2^{(2)},S_1^{(3)},S_2^{(3)}}} 
E_4^{*}E_3E_2^{*}E_1 e^{i\phi}
\\  &
\mu_{S_1^{(0)}S_2^{(0)}}\mu_{S_2^{(1)}S_1^{(1)}}
\mu_{S_1^{(2)}S_2^{(2)}}\mu_{S_2^{(3)}S_1^{(3)}}
\zeta_{S_1^{(0)}}
2\pi\delta(\omega_1-\omega_2+\omega_3-\omega_4)
\int_{-\infty}^{+\infty}\frac{d\epsilon_1}{2\pi}\int_{-\infty}^{+\infty}\frac{d\epsilon_2}{2\pi}
e^{iT_2(\omega_1-\omega_2+\epsilon_1-\epsilon_2)}
\nonumber 
\\ &\qquad % 1
\bigg(
G^{<}_{S_1^{(1)}S_1^{(0)}}(\epsilon_1) G^{>}_{S_2^{(0)}S_2^{(3)}}(\epsilon_1+\Omega_3+\omega_4)
G^{>}_{S_1^{(3)}S_1^{(2)}}(\epsilon_2) G^{>}_{S_2^{(2)}S_2^{(1)}}(\epsilon_1+\Omega_1+\omega_1)
\nonumber 
\\ &\qquad + % 2
G^{<}_{S_1^{(1)}S_1^{(2)}}(\epsilon_2-\Omega_1-\omega_1) G^{>}_{S_2^{(2)}S_2^{(3)}}(\epsilon_1)
G^{>}_{S_1^{(3)}S_1^{(0)}}(\epsilon_2-\Omega_3-\omega_4) G^{>}_{S_2^{(0)}S_2^{(1)}}(\epsilon_2)
\nonumber 
\\ &\qquad + % 3
G^{<}_{S_1^{(3)}S_2^{(1)}}(\epsilon_2) G^{>}_{S_1^{(1)}S_1^{(2)}}(\epsilon_2-\Omega_1-\omega_1)
G^{>}_{S_2^{(2)}S_1^{(0)}}(\epsilon_1) G^{>}_{S_2^{(0)}S_2^{(3)}}(\epsilon_1+\Omega_3+\omega_4)
\nonumber 
\\ &\qquad + % 4
G^{<}_{S_2^{(2)}S_2^{(1)}}(\epsilon_1+\Omega_1+\omega_1) G^{>}_{S_1^{(1)}S_2^{(3)}}(\epsilon_1)
G^{>}_{S_1^{(3)}S_1^{(0)}}(\epsilon_2-\Omega_3-\omega_4) G^{>}_{S_2^{(0)}S_1^{(2)}}(\epsilon_2)
\nonumber 
\\ &\qquad - % 5
G^{<}_{S_2^{(2)}S_2^{(1)}}(\epsilon_1+\Omega_1+\omega_1) G^{>}_{S_1^{(1)}S_1^{(0)}}(\epsilon_1)
G^{>}_{S_2^{(0)}S_2^{(3)}}(\epsilon_1+\Omega_3+\omega_4) G^{>}_{S_1^{(3)}S_1^{(2)}}(\epsilon_2)
\nonumber 
\\ &\qquad - % 6
G^{<}_{S_1^{(1)}S_2^{(3)}}(\epsilon_1) G^{>}_{S_1^{(3)}S_1^{(0)}}(\epsilon_2-\Omega_3-\omega_4)
G^{>}_{S_2^{(0)}S_1^{(2)}}(\epsilon_2) G^{>}_{S_2^{(2)}S_2^{(1)}}(\epsilon_1+\Omega_1+\omega_1)
\nonumber 
\\ &\qquad - % 7
G^{<}_{S_1^{(1)}S_1^{(2)}}(\epsilon_2-\Omega_1-\omega_1) G^{>}_{S_2^{(2)}S_1^{(0)}}(\epsilon_1)
G^{>}_{S_2^{(0)}S_2^{(3)}}(\epsilon_1+\Omega_3+\omega_4) G^{>}_{S_1^{(3)}S_2^{(1)}}(\epsilon_2)
\nonumber 
\\ &\qquad - % 8
G^{<}_{S_2^{(0)}S_2^{(1)}}(\epsilon_2) G^{>}_{S_1^{(1)}S_1^{(2)}}(\epsilon_2-\Omega_1-\omega_1)
G^{>}_{S_2^{(2)}S_2^{(3)}}(\epsilon_1) G^{>}_{S_1^{(3)}S_1^{(0)}}(\epsilon_2-\Omega_3-\omega_4)
\bigg)
\nonumber 
\end{align}
\end{widetext}
This result is alternative to Eq.(10) in Ref.~\cite{MukamelJCP15}
theoretical description of 2D optical spectroscopy in junctions.
While no experimental result on multi-dimensional optical spectroscopy in junctions
have been reported yet, first proposals on utilizing pump-probe approaches
for junctions diagnostics were reported recently~\cite{SelzerPeskinJPCC13,OchoaSelzerPeskinMGJPCL15}. 

Contrary to the usually employed QME based considerations of optical response 
Eq.~(\ref{SstimOmT}) avoids assumption of molecule-contacts destruction of 
coherence at times of interaction with radiation field.
To demonstrate advantage of the formulation we consider a simple model of molecular
junction (see Fig.~\ref{fig4}) with two low-lying orbitals (e.g., HOMO and HOMO-1)
hybridized with  states of contacts and (through the contacts) with each other and 
one higher lying orbital (e.g., LUMO). The model is described by Hamiltonian 
(\ref{H})-(\ref{VMF}) with eight many-body molecular states $\{\lvert S\rangle\}$
\begin{equation}
\begin{split}
& 
\lvert S_1\rangle=\lvert 0,0,0\rangle, 
\\ &
\lvert S_2\rangle=\lvert 1,0,0\rangle, \,
\lvert S_3\rangle=\lvert 0,1,0\rangle, \,
\lvert S_4\rangle=\lvert 0,0,1\rangle, 
\\& 
\lvert S_5\rangle=\lvert 0,1,1\rangle, \,
\lvert S_6\rangle=\lvert 1,0,1\rangle,  \,
\lvert S_7\rangle=\lvert 1,1,0\rangle,
\\ &
\lvert S_8\rangle=\lvert 1,1,1\rangle 
\end{split}
\end{equation}
Energies of the states are $0$, $\varepsilon_1$, $\varepsilon_2$, $\varepsilon_3$,
$\varepsilon_2+\varepsilon_3$, $\varepsilon_1+\varepsilon_3$, 
$\varepsilon_1+\varepsilon_2$, and $\varepsilon_1+\varepsilon_2+\varepsilon_3$,
respectively. 
Quasiparticle representation of the model Hamiltonians for the molecule and its coupling
with contacts is
\begin{align}
\hat H_M=& \sum_{m=1}^3\varepsilon_m\hat d_m^\dagger\hat d_m
\\
\hat V_{MK} =& \sum_{k\in K}\sum_{m=1}^3\bigg(V_{km}\hat c_k^\dagger\hat d_m+H.c.\bigg) 
\end{align}
Here $\hat d_m^\dagger$ ($\hat d_m$) creates (annihilates) electron in orbital $m$.
Hybridization of molecular orbitals with states in the contacts is characterized by
dissipation matrix ($K=L,R$)
\begin{equation}
\label{GammaK}
\Gamma^K_{m_1m_2}(E) = 2\pi\sum_{k\in K} V_{m_1k}V_{km_2}\delta(E-\varepsilon_k)
\end{equation}
which in the wide-band approximation is assumed energy independent.
Note that for the model of Fig.~\ref{fig4} we assume contact induced
hybridization between orbitals $1$ and $2$, while orbital $3$ does not
hybridize with the low lying levels. The model is reasonable,
because usual HOMO-LUMO gaps in molecules are at least of the order of $1$~eV,
which makes hybridization between HOMOs and LUMOs through the contacts negligible. 

%%%%%%%%%%%%%%%%%%%%%%%%%%%%%%%%%%%%%%%%%%%%%%
\begin{figure}[htpb]
\centering\includegraphics[width=0.7\linewidth]{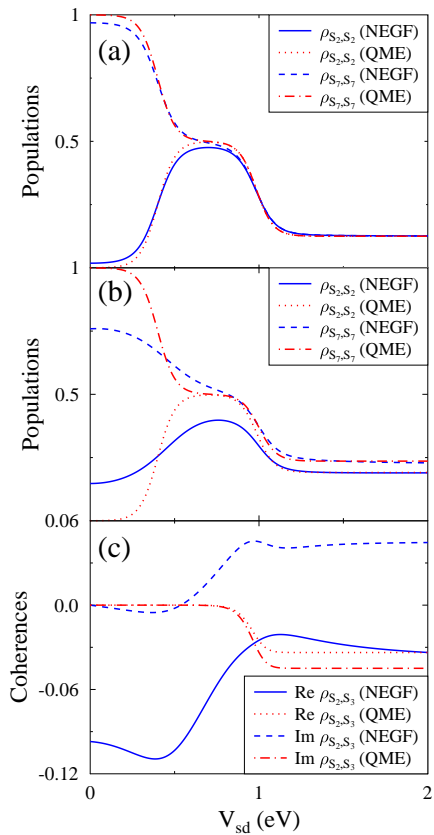}
\caption{\label{fig5}
Reduced density matrix for the model of Fig.~\ref{fig4} vs. applied bias.
Redfield QME results are compared to exact NEGF results for
(a) weak diagonal and (b) strong non-diagonal molecule-contacts couplings.
See text for parameters. 
}
\end{figure}
%%%%%%%%%%%%%%%%%%%%%%%%%%%%%%%%%%%%%%%%%%%%%%
%%%%%%%%%%%%%%%%%%%%%%%%%%%%%%%%%%%%%%%%%%%%%%
\begin{figure*}[htpb]
%\centering\includegraphics[width=0.75\linewidth,angle=90]{fig6}
\centering\includegraphics[width=\linewidth]{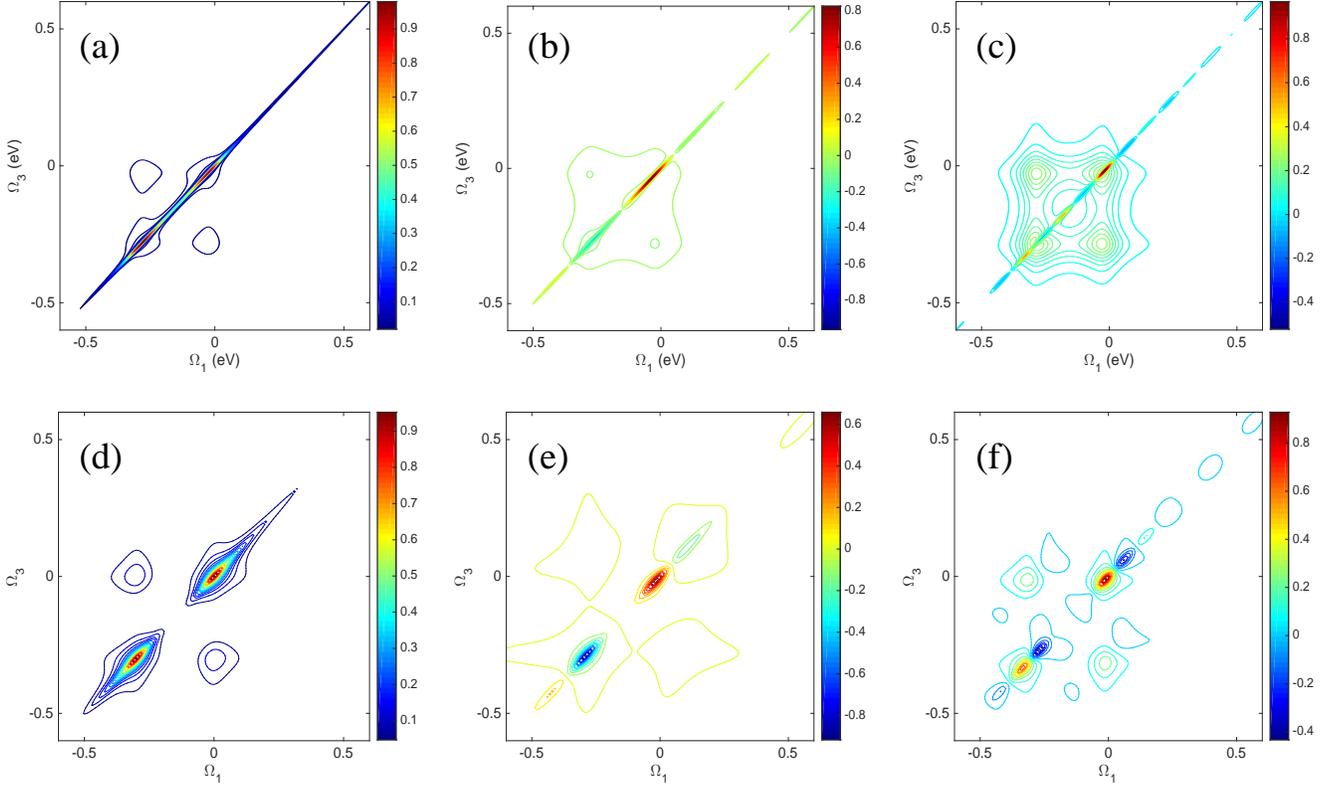}
\caption{\label{fig6}
The stimulated signal $S_{stim}(\Omega_3,T_2,\Omega_1)$ for the model
of Fig.~\ref{fig4} at equilibrium, $V_{sd}=0$. 
Approximate PP-NEGF results, Eq.~(\ref{SstimOmT}), 
are compared with exact NEGF calculations, Eq.~(\ref{SstimNEGF}).
Results for delay times $T_2=0$, $10$ and $25$~fs are shown
in panels (a)-(c) for PP-NEGF and (d)-(f) for NEGF.
Maximum of the signal is scaled to $1$.
See text for parameters. 
}
\end{figure*}
%%%%%%%%%%%%%%%%%%%%%%%%%%%%%%%%%%%%%%%%%%%%%%

For the model of Fig.~\ref{fig4} the QME approach will fail to predict
stimulated signal (\ref{SstimOmT}). From the QME expression for the 
$S_{stim}(\Omega_1,T_2,\Omega_3)$, Eq.~(\ref{SstimQME}),
one sees that coupling (in the absence of the field) between states characterizing 
optical transition (pair $S_1^{(0)}S_2^{(0)}$ in Eq.~(\ref{SstimQME});
pairs of states ($\lvert S_2\rangle,\lvert S_4\rangle$), 
($\lvert S_3\rangle,\lvert S_4\rangle$), 
($\lvert S_7\rangle,\lvert S_5\rangle$) and
($\lvert S_7\rangle,\lvert S_6\rangle$) of the model)
is necessary for non-zero stimulated signal.
In the absence of optical field such coupling cam be provided only by
contacts, i.e. by dissipation matrix $\Gamma$, Eq.~(\ref{GammaK}).
As discussed above, for states separated by $\sim 1$~eV such coupling is negligible.
At the same time in the presence of such coupling (for example, for
closer lying pairs of states) accurate calculation of the reduced density matrix,
which is part of the expression (\ref{SstimQME}), is required for proper prediction
of the signal. At the usual level of consideration (second order in the system-bath 
coupling) QME is known to fail in this regime~\cite{EspMGJPCC10}. 
Fig.~\ref{fig5} compares Redfield QME simulations of the reduced density matrix
to NEGF results. The latter are exact for the model of Fig.~\ref{fig4}.
Parameters of the simulations are $T=300$~K, $\varepsilon_1=-0.5$~eV,
$\varepsilon_2=-0.2$~eV, $\varepsilon_3=0.5$~eV.
Fermi energy is taken as origin, $E_F=0$, and bias is applied symmetrically,
$\mu_L=E_F+|e|V_{sd}/2$ and $\mu_R=E_F-|e|V_{sd}/2$.
Results of simulations presented in Fig.~\ref{fig5}a employed weak
diagonal coupling between molecule and contacts,
$\Gamma^K+{m_1,m_2}=\delta_{m_1,m_2}\, 0.01$~eV ($m_{1,2}\in\{1,2,3\}$, $K=L,R$).
Calculations in Figs.~\ref{fig5}b and \ref{fig5}c utilized stronger non-diagonal coupling
parameters for levels $1$ and $2$: $\Gamma^L_{11}=0.2$~eV, $\Gamma^R_{11}=0.05$~eV,
$\Gamma^K_{12}=\Gamma^K_{21}=\Gamma^K_{22}=0.1$~eV ($K=L,R$).
Fig.~\ref{fig5}a shows that for weak diagonal coupling Redfield QME is pretty accurate.
However, as shown in Figs.~\ref{fig5}b and \ref{fig5}c, 
in the presence of non-diagonal coupling (this is the situation necessary
for simulations of the optical signal within the QME approach) Redfield QME
predictions deviate significantly from exact NEGF results 
(especially in prediction of coherences).
  
 %%%%%%%%%%%%%%%%%%%%%%%%%%%%%%%%%%%%%%%%%%%%%%
\begin{figure*}[htpb]
\centering\includegraphics[width=\linewidth]{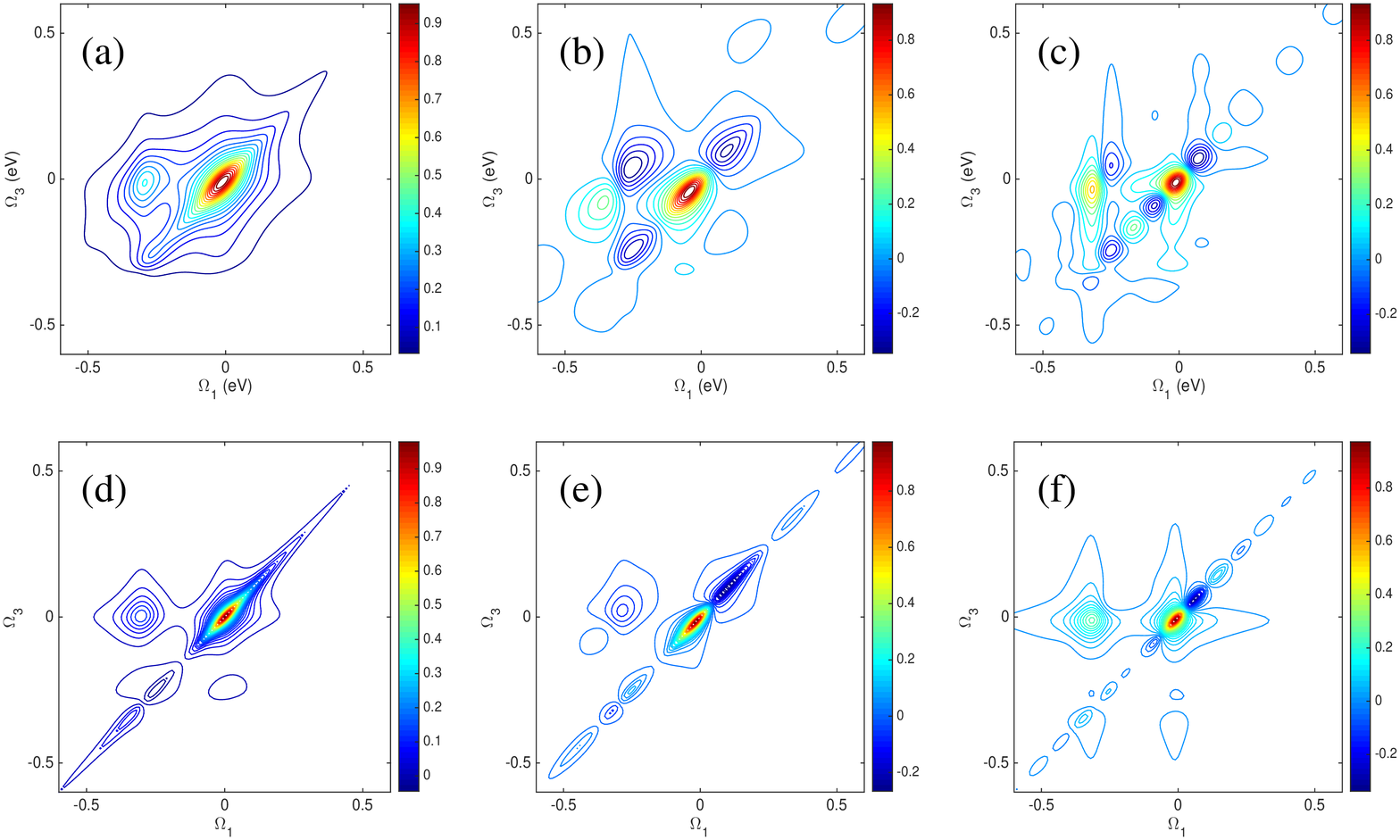}
\caption{\label{fig7}
The stimulated signal $S_{stim}(\Omega_3,T_2,\Omega_1)$ for the model
of Fig.~\ref{fig4} in biased junction, $V_{sd}=2$~V. 
Approximate PP-NEGF results, Eq.~(\ref{SstimOmT}), 
are compared with exact NEGF calculations, Eq.~(\ref{SstimNEGF}).
Results for delay times $T_2=0$, $10$ and $25$~fs are shown
in panels (a)-(c) for PP-NEGF and (d)-(f) for NEGF.
Maximum of the signal is scaled to $1$.
See text for parameters. 
}
\end{figure*}
%%%%%%%%%%%%%%%%%%%%%%%%%%%%%%%%%%%%%%%%%%%%%%

Contrary to the QME approximate PP-NEGF expression, Eq.~(\ref{SstimOmT}),
is capable to reproduce stimulated signal for the model of Fig.~\ref{fig4}.
Figure~\ref{fig6} compares results of simulations utilizing approximate PP-NEGF 
expression (\ref{SstimOmT}) with exact NEGF results, Eq.~(\ref{SstimNEGF}).
Parameters of the simulations are $\Gamma^K_{m_1,m_2}=0.05$~eV 
($m_{1,2}=1,2$, $K=L,R$), $\omega_1=\omega_4=1$~eV, $\omega_2=\omega_3=0.7$~eV.
Other parameters are as in Fig.~\ref{fig5}.
Peak at $\Omega_1=\Omega_3=0$ represents the $1\to 3$ transiiton
($\lvert S_1\rangle to \lvert S_4\rangle$ and $\lvert S_7\rangle\to \lvert S_5\rangle$ 
in terms of states), $\Omega_1=\Omega_3=-0.3$~eV represents
transition $2\to 3$ ($\lvert S_3\rangle\to\lvert S_4\rangle$ and
$\lvert S_7\to \lvert S_6\rangle$ in terms of states).
Off-diagonal peaks ($\Omega_1=0$, $\Omega_3=-0.3$~eV 
and  $\Omega_1=-0.3$~eV, $\Omega_3=0$) indicate correlation between
levels $1$ and $3$, pairs of states ($\lvert S_2\rangle$,$\lvert S_3\rangle$)
and ($\lvert S_5\rangle$,$\lvert S_6\rangle$) due to coupling to contacts.
One sees that main features of the spectrum are reproduced qualitatively correctly.

Figure~\ref{fig7} shows stimulated signal in a biased junction. 
Parameters of the simulations are $\Gamma^L_{11}=0.1$~eV,
$\Gamma^R_{11}=\Gamma^R_{12}=\Gamma^R_{21}=0$.
Other parameters are as in Fig.~\ref{fig6}.
Here level $1$ is attached to the left contact only.
Also here stimulated signal is reproduced qualitatively correctly
by the approximate expression (\ref{SstimOmT}).
It is interesting to note that under bias the signal can provide information on asymmetry
in the junction. At high bias of $V_{sd}=2$~V levels $2$ and $3$ have population of 
$1/2$, while level $1$ is fully populated. As a result only transition $1\to 3$
is seen in the signal, while electronic transitions in the $2\to 3$ channel are compensated
by electronic transition in $3\to 2$ channel (or hole transitions in $2\to 3$).
This is easily seen from the structure of Eq.~(\ref{SstimNEGF}).
As a result only one diagonal and one off-diagonal peaks are visible.

%%%%%%%%%%%%%%%%%%%%%%%%%%%%%%%%%%%%%%%%%%%%%%

\section{Conclusion}\label{conclude}
We consider simulation of optical response functions in molecular junctions.
Following standard methodology of the nonlinear optical spectroscopy
and restricting consideration to classical radiation fields we first express
optical response functions in the form convenient for implementation of Green function 
techniques, and then propose a simple approximate scheme for representing
the response functions in terms of the pseudoparticle nonequilibrium Green functions
(PP-NEGF). Similar to more traditional quantum master equation based approach,
our formulation is capable to describe optical response in the basis of many-body
states of the system.  It also accounts approximately for hybridization 
of molecular states due to coupling to contacts. Finally, it avoids approximation
of the standard QME treatment when each instant of interaction with light
results in destruction of molecule-contacts entanglement.  
Within simple 3-level model (e.g., HOMO-1, HOMO and LUMO)
and utilizing stimulated signal in the fourth order of optical field we illustrate 
the advantages of the proposed approximate scheme.
Comparing results of the PP-NEGF simulations with exact (for the chosen noninteracting
model) NEGF results we show that the formulation reproduces stimulated signal
qualitatively correctly, when QME based approach fails.
Finally, we show that 2d stimulated signal can provide information on asymmetry in
junctions.

%%%%%%%%%%%%%%%%%%%%%%%%%%%%%%%%%%%%%%%%%%%%%%
%%%%%%%%%%%%%%%%%%%%%%%%%%%%%%%%%%%%%%%%%%%%%%

\begin{acknowledgments}
M.G. gratefully acknowledges support by the US Department of Energy
(Early Career Award, DE-SC0006422).
\end{acknowledgments}

%%%%%%%%%%%%%%%%%%%%%%%%%%%%%%%%%%%%%%%%%%%

\appendix
\section{QME expression for stimulated signal}\label{appA}
Employing quantum regression theorem in evaluation of multi-time correlation
functions in Eq.(\ref{S3c}) leads to the following expression for stimulated
signal at steady-state transport (compare with Eq.(10) of Ref.~\cite{MukamelJCP15})
\begin{widetext}
\begin{align}
\label{SstimQME}
& S_{stim}(\Omega_1,T_2,\Omega_3)\equiv 
\int_0^\infty dT_3\int_0^\infty dT_1 e^{i\Omega_3T_3+i\Omega_1 T_1}
S_{stim}(T_3,T_2,T_1)
\\ &
= 2\,\mbox{Im}\sum_{\substack{S_1^{(0)},S_2^{(0)},S_1^{(1)},S_2^{(1)}\\
                                                   S_1^{(2)},S_2^{(2)},S_1^{(3)},S_2^{(3)}}} 
 \sum_{S_a,S_b,S_c} E_4^{*}E_3E_2^{*}E_1 e^{iT_2(\omega_1-\omega_2)+i\phi}
\nonumber \\  &
\mu_{S_1^{(0)}S_2^{(0)}}\mu_{S_2^{(1)}S_1^{(1)}}
\mu_{S_1^{(2)}S_2^{(2)}}\mu_{S_2^{(3)}S_1^{(3)}}\,
2\pi\delta(\omega_1-\omega_2+\omega_3-\omega_4)
\nonumber 
\\ &\qquad % 1
\bigg(\ \
\mathcal{G}^{r}_{S_1^{(0)}S_2^{(0)},S_cS_2^{(3)}}(\Omega_3+\omega_4)\, 
\mathcal{G}^{r}_{S_cS_1^{(3)},S_bS_1^{(2)}}(T_2)\,
\mathcal{G}^{r}_{S_bS_2^{(2)},S_aS_2^{(1)}}(\Omega_1+\omega_1)\,
\rho_{S_1^{(1)}S_a}
\nonumber 
\\ &\qquad - % 2
\mathcal{G}^{r}_{S_1^{(0)}S_2^{(0)},S_1^{(3)}S_c}(\Omega_3+\omega_4)\, 
\mathcal{G}^{r}_{S_2^{(3)}S_c,S_2^{(2)}S_b}(T_2)\
\mathcal{G}^{r}_{S_1^{(2)}S_b,S_1^{(1)}S_a}(\Omega_1+\omega_1)\,
\rho_{S_aS_2^{(1)}}
\nonumber 
\\ &\qquad + % 3
\mathcal{G}^{r}_{S_1^{(0)}S_2^{(0)},S_1^{(3)}S_c}(\Omega_3+\omega_4)\, 
\mathcal{G}^{r}_{S_2^{(3)}S_c,S_2^{(2)}S_b}(T_2)\,
\mathcal{G}^{r}_{S_1^{(2)}S_b,S_aS_2^{(1)}}(\Omega_1+\omega_1)\,
\rho_{S_1^{(1)}S_a}
\nonumber 
\\ &\qquad - % 4
\mathcal{G}^{r}_{S_1^{(0)}S_2^{(0)},S_cS_2^{(3)}}(\Omega_3+\omega_4)\, 
\mathcal{G}^{r}_{S_cS_1^{(3)},S_bS_1^{(2)}}(T_2)\,
\mathcal{G}^{r}_{S_bS_2^{(2)},S_1^{(1)}S_a}(\Omega_1+\omega_1)\, 
\rho_{S_aS_2^{(1)}}
\nonumber 
\\ &\qquad + % 5
\mathcal{G}^{r}_{S_1^{(0)}S_2^{(0)},S_cS_2^{(3)}}(\Omega_3+\omega_4)\, 
\mathcal{G}^{r}_{S_cS_1^{(3)},S_2^{(2)}S_b}(T_2)\,
\mathcal{G}^{r}_{S_1^{(2)}S_b,S_1^{(1)}S_a}(\Omega_1+\omega_1)\,
\rho_{S_aS_2^{(1)}}
\nonumber 
\\ &\qquad - % 6
\mathcal{G}^{r}_{S_1^{(0)}S_2^{(0)},S_1^{(3)}S_c}(\Omega_3+\omega_4)\,
\mathcal{G}^{r}_{S_2^{(3)}S_c,S_bS_1^{(2)}}(T_2)\,
\mathcal{G}^{r}_{S_bS_2^{(2)},S_aS_2^{(1)}}(\Omega_1+\omega_1)\,
\rho_{S_1^{(1)}S_a}
\nonumber 
\\ &\qquad + % 7
\mathcal{G}^{r}_{S_1^{(0)}S_2^{(0)},S_1^{(3)}S_c}(\Omega_3+\omega_4)\,
\mathcal{G}^{r}_{S_2^{(3)}S_c,S_bS_1^{(2)}}(T_2)\,
\mathcal{G}^{r}_{S_bS_2^{(2)},S_1^{(1)}S_a}(\Omega_1+\omega_1)\,
\rho_{S_aS_2^{(1)}}
\nonumber 
\\ &\qquad - % 8
\mathcal{G}^{r}_{S_1^{(0)}S_2^{(0)},S_cS_2^{(3)}}(\Omega_3+\omega_4)\, 
\mathcal{G}^{r}_{S_cS_1^{(3)},S_2^{(2)}S_b}(T_2)\,
\mathcal{G}^{r}_{S_1^{(2)}S_b,S_aS_2^{(1)}}(\Omega_1+\omega_1)\, 
\rho_{S_1^{(1)}S_a}\ \
\bigg)
\nonumber 
\end{align}
\end{widetext}
Here $\rho_{ab}$ is the reduced density matrix,
\begin{equation}
\mathcal{G}^r_{ab,cd}(t)\equiv -i\theta(t)\ll ba\rvert e^{-i\mathcal{L}t}\lvert dc\gg
\end{equation}
is the Liouville space retarded Green function~\cite{EspMGPRB09},
and $\mathcal{L}$ is the Liouvillian.
Fourier transform of the retarded Green function is
\begin{equation}
\mathcal{G}^r_{ab,cd}(E) = \sum_\gamma
\frac{\ll ba\vert R_\gamma\gg\,\ll L_\gamma\vert dc\gg}{E-\lambda_\gamma+i\delta}
\end{equation} 
where $\delta\to 0^{+}$ and $\lambda_\gamma$, $\lvert R_\gamma\gg$
and $\ll L_\gamma\rvert$ are eigenvalues and right and left eigenvectors of 
the Liouvillian, respectively
\begin{equation}
\mathcal{L} = \sum_\gamma \lvert R_\gamma\gg\, \lambda_\gamma\,
\ll L_\gamma\rvert
\end{equation} 

%%%%%%%%%%%%%%%%%%%%%%%%%%%%%%%%%%%%%%%%%%%

\section{NEGF expression for stimulated signal}\label{appB}
For the quadratic (non-interacting) model of of Fig.~\ref{fig4}
multi-time correlation functions in Eq.(\ref{S3c}) can be evaluated
employing the Wick's theorem~\cite{FetterWalecka_1971}.
This leads to exact expression for stimulated signal 
\begin{widetext}
\begin{align}
\label{SstimNEGF}
& S_{stim}(\Omega_1,T_2,\Omega_3)\equiv 
\int_0^\infty dT_3\int_0^\infty dT_1 e^{i\Omega_3T_3+i\Omega_1 T_1}
S_{stim}(T_3,T_2,T_1)
=2\,\mbox{Re}\sum_{\substack{m_0,m_1\\m_2,m_3}=\{1,2\}} 
E_4^{*}E_3E_2^{*}E_1\, e^{i\phi}
\nonumber \\  &
\mu_{m_03}\mu_{3m_1}\mu_{m_23}\mu_{3m_3}\,
2\pi\delta(\omega_1-\omega_2+\omega_3-\omega_4)
\int_{-\infty}^{+\infty}\frac{d\epsilon_1}{2\pi}\int_{-\infty}^{+\infty}\frac{d\epsilon_2}{2\pi}\,
e^{iT_2(\omega_1-\omega_2+\epsilon_1-\epsilon_2)}
\nonumber \\ &
\bigg(\ \,\bigg[G^{<}_{m_1m_2}(\epsilon_2-\omega_1-\omega_1)\, G_{33}^{>}(\epsilon_2)
-G^{>}_{m_1m_2}(\epsilon_2-\Omega_1-\omega_1)\, G_{33}^{<}(\epsilon_2)\bigg]
A_{m_3m_0}(\epsilon_2-\Omega_3-\omega_4)\, A_{33}(\epsilon_1)
\nonumber \\ &
+\bigg[G_{m_1m_0}^{<}(\epsilon_1)\, G_{33}^{>}(\epsilon_1+\Omega_1+\omega_1)
-G_{m_1m_0}^{>}(\epsilon_1)\, G_{33}^{<}(\epsilon_1+\Omega_1+\omega_1)\bigg]
A_{m_3m_2}(\epsilon_2)\, A_{33}(\epsilon_1+\Omega_3+\omega_4)\bigg)
\end{align}
\end{widetext}
Here $G^{<(>)}_{m_1m_2}(E)$ is Fourier transform of 
the lesser (greater) projection of the quasiparticle Green function
\begin{equation}
 G_{m_1m_2}(\tau_1,\tau_2) \equiv -i\langle T_c\,
 \hat d_{m_1}(\tau_1)\,\hat d_{m_2}^\dagger(\tau_2)\rangle
\end{equation}
and $A_{m_1m_2}(E)\equiv i[G^{>}_{m_1m_2}(E)-G^{<}_{m_1m_2}(E)]$
is quasiparticle spectral function.

%%%%%%%%%%%%%%%%%%%%%%%%%%%%%%%%%%%%%%%%%%%
%\bibliography{pt_mol}

%merlin.mbs apsrev4-1.bst 2010-07-25 4.21a (PWD, AO, DPC) hacked
%Control: key (0)
%Control: author (8) initials jnrlst
%Control: editor formatted (1) identically to author
%Control: production of article title (-1) disabled
%Control: page (0) single
%Control: year (1) truncated
%Control: production of eprint (0) enabled
%

\end{document}